\documentclass[journal]{vgtc}                     
\usepackage{mathptmx}       
\usepackage{graphicx}       
\usepackage{xcolor}         
\usepackage{colortbl}       
\usepackage{tabularx}       
\usepackage{tabu}           
\usepackage{booktabs}       
\usepackage{lipsum}         
\usepackage{mwe}            
\usepackage{float}
\usepackage{multirow}

\usepackage{tikz}
\usepackage{forest}
\usetikzlibrary{positioning}
\usepackage[subtle]{savetrees}

\graphicspath{{figs/}{figures/}{pictures/}{images/}{./}}

\title{UnrealVis: A Testing Laboratory of Optimization Techniques in Unreal Engine
for Scientific Visualization}

\author{%
  \authororcid{Matteo Filosa}{0009-0004-8868-7907},
  Andrea Nardocci, 
  \authororcid{Tiziana Catarci}{0000-0002-3578-1121}, and
  \authororcid{Marco Angelini}{0000-0001-9051-6972}
}

\authorfooter{
  \item Matteo Filosa, Andrea Nardocci, and Tiziana Catarci are with Sapienza University of Rome, Italy. E-mail: \{m.filosa, angelini, catarci\}@diag.uniroma1.it, nardocci.1709323@studenti.uniroma1.it
  \item Tiziana Catarci is with the Institute for Cognitive Sciences and Technologies (ISTC) -National Research Council (Cnr), in Rome, Italy
  \item {Marco Angelini is with Link Campus University of Rome, Italy: E-mail: m.angelini@unilink.it}
}

\abstract{
Visualizing large 3D scientific datasets requires balancing performance and fidelity, but traditional tools often demand excessive technical expertise. We introduce UnrealVis, an Unreal Engine optimization laboratory for configuring and evaluating rendering techniques during interactive exploration. Following a review of 55 papers, we established a taxonomy of 22 optimization techniques across six families, implementing them through engine subsystems such as Nanite, Level of Detail(LOD) schemes, and culling. The system features an intuitive workflow with live telemetry and A/B comparisons for local and global performance analysis. Validated through case studies of ribosomal structures and volumetric flow fields, along with an expert evaluation, UnrealVis facilitates the selection of optimization combinations that meet performance goals while preserving structural fidelity. UnrealVis is available at \url{https://github.com/XAIber-lab/UnrealVis}}


\teaser{
  \centering
  \includegraphics[width=1.0\linewidth, alt={The UnrealVis graphical user interface displaying a dense molecular structure in the center, a live performance line chart with a two-way whisker on the left, and a helper control overlay on the right.}]{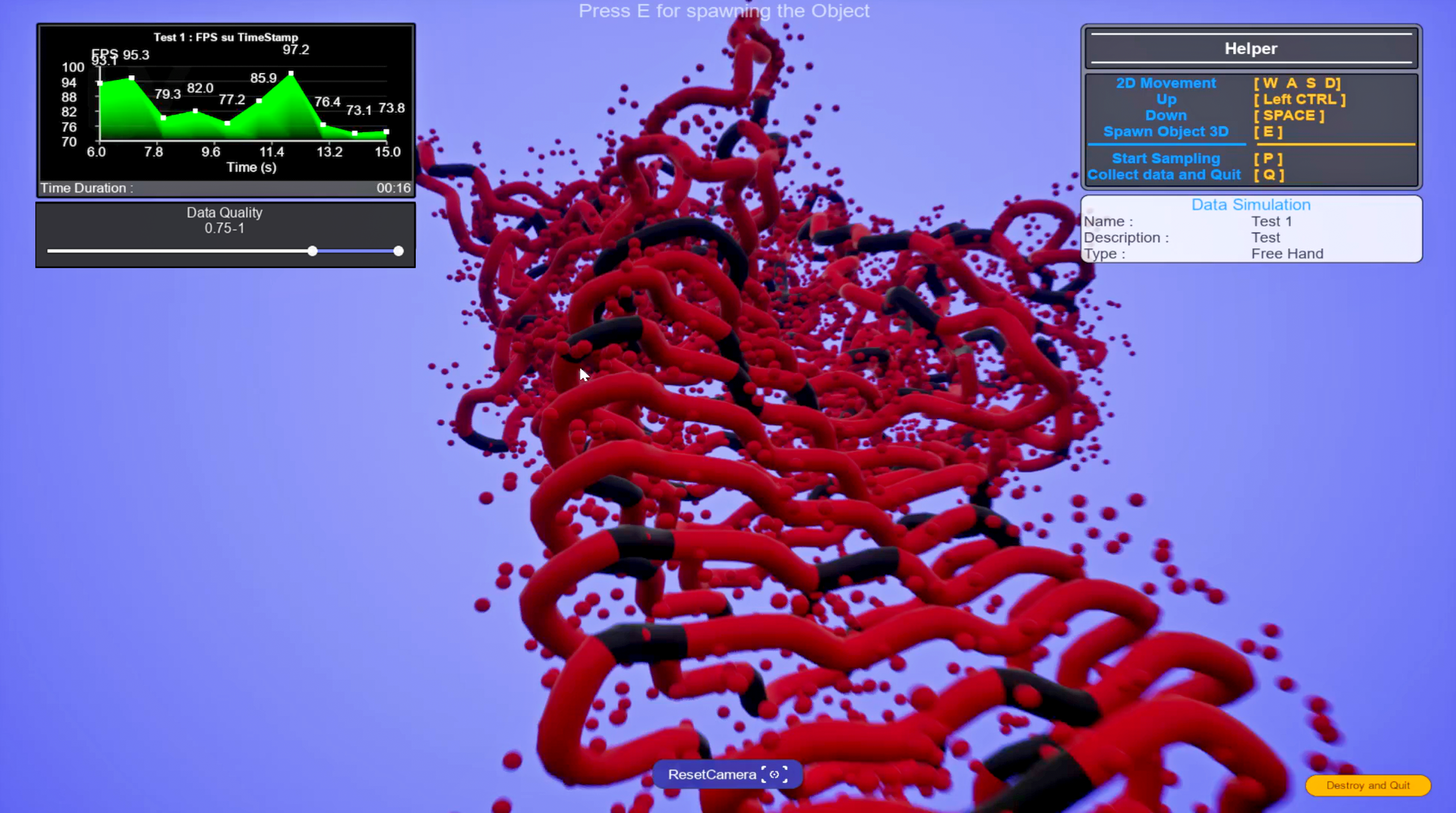}
  \caption{Interactive visualization of the 3SYJ adhesin dataset \cite{pdb_3syj, Berman2000PDB, Meng2011} within UnrealVis. The molecular model is rendered as a dense ribbon-and-sphere 3D point cloud to stress-test navigation at high geometric density. The interface features a performance panel with a two-way whisker tool for local stability analysis (left) and a helper overlay for interaction controls and simulation configuration (right).}
  \label{fig:teaser}
}

\renewcommand{\manuscriptnotetxt}{}
\begin{document}

\maketitle

\section{Introduction}
\label{sec:intro}

In the contemporary landscape of digital technology, the visualization of complex, high-dimensional data has emerged as a fundamental challenge across myriad scientific domains, from molecular biology to fluid dynamics. Scientific visualization focuses on developing graphical methods that allow researchers to extract underlying models of particular phenomena~\cite{reina2020mtv}. As datasets continue to grow in scale and complexity, reaching the petabyte range in domains such as combustion simulation, the need to select and implement appropriate visualization techniques becomes increasingly critical~\cite{Ayachit2015}. Effective exploration is currently hindered by recurring bottlenecks: \textit{Big Data} poses severe challenges in memory processing; \textit{data representation} requires sophisticated geometric abstractions to depict complex structures accurately; and \textit{fluidity in interactivity} is often compromised when high visual fidelity is demanded, especially in immersive environments~\cite{drouhard2015immersive,lanrezac2021wielding}. Historically, the interactive manipulation of such large-volume data was constrained by significant hardware limitations, leading to a reliance on private high-performance computing (HPC) solutions and specialized, often proprietary software frameworks~\cite{Ingo2016,wang2023lvdif}. While these tools provide the necessary rendering power, they can be exclusive, creating a barrier between data generation and intuitive exploration~\cite{reina2020mtv}. In the present era, however, the advent of robust game engines has opened new avenues for high-fidelity visualization. Originally engineered for interactive entertainment, platforms like Unreal Engine 5 (UE5) have matured into sophisticated rendering environments. UE5 introduces virtualized geometry systems (Nanite) and dynamic global illumination models (Lumen) that offer unprecedented capabilities for handling complex datasets previously unrenderable in real-time~\cite{marsden2020unreal,EpicNanite2021,EpicLumen2021}. However, transitioning scientific visualization into a game engine environment is not without complications. Researchers often encounter scenarios where simply importing data into a game engine does not yield immediate performance profits, or where the sheer volume of data overwhelms standard engine buffers~\cite{wang2023lvdif}. Crucially, while game engines offer a plethora of optimization techniques (e.g., culling, instancing), the scientific visualization community lacks a systematic environment for isolating, testing, and quantifying the impact of these techniques across diverse datasets. Optimization is often handled in an ad hoc, hardcoded manner, making it difficult to balance raw rendering speed with the strict visual fidelity required for scientific analysis. This research addresses the critical need for an empirical, reproducible approach to performance tuning in game engine-driven scientific visualization. Rather than presenting a bespoke, domain-specific viewer, we introduce a diagnostic laboratory that decouples rendering optimization from data ingestion, enabling researchers to systematically evaluate how different techniques affect both telemetry and visual accuracy. In summary, we contribute: 
\begin{itemize}
\item a systematic literature review of state-of-the-art optimizations in 3D scientific visualization, leading to a comprehensive taxonomy of techniques adaptable to modern game engines;
\item \textit{UnrealVis}, an interactive testing framework natively built in UE5 to implement, monitor, and compare various optimization profiles across heterogeneous scientific datasets (specifically, biomolecular structures and turbulent flow fields);
\item a multi-tiered evaluation comprising case studies, a formative expert user study, an automated benchmarking suite based on deterministic camera paths to provide quantitative analysis helping in identifying performance bottlenecks in game-engine-driven scientific visualization.
\end{itemize}
UnrealVis is available at \url{https://github.com/XAIber-lab/UnrealVis}

\section{Related Work}
\label{sec:rel}

Scientific visualization has traditionally relied on specialized desktop frameworks such as ParaView, VMD, PyMOL, and OSPRay-based systems, which offer scalable rendering pipelines and domain-specific interaction models for large scientific datasets~\cite{Ayachit2015,VMD1996,Warren2012,Ingo2016}. These tools provide robust support for parallel rendering, out-of-core data management, and scripting, but they often require complex setup and lack the tight integration with modern real-time rendering and interaction paradigms that game engines offer. 
Over the last decade, several research efforts have explored the use of game engines as flexible back-ends for scientific visualization. Unity-based molecular viewers, such as UnityMol, have demonstrated how GPU shaders and techniques like HyperBalls can be exploited to render large biomolecular structures interactively, supporting formats such as Protein Data Bank (PDB), mmCIF, and trajectory files while targeting Virtual Reality (VR) and desktop environments alike~\cite{UnityMol2013,Chavent2011,Laureanti2020}. Similar approaches apply GPU instancing and custom shading to handle tens of thousands of atoms at interactive frame rates, enabling dynamic operations such as cross-section cutting and real-time molecular analysis~\cite{LeMuzic2015}. Unreal Engine has likewise been adopted to visualize a diverse range of phenomena, including cosmological volumes, seismic or earthquake simulations, radiation data, molecular datasets, and even holographic rendering of complex scenes~\cite{marsden2020unreal,Kreylos2012,Greenwood2020,choi2024chimera}. In these works, authors leverage built-in capabilities such as high-quality physically based rendering, virtualized geometry (Nanite), dynamic global illumination (Lumen), and engine-level culling and streaming to maintain real-time interaction on complex scenes~\cite{EpicNanite2021,EpicLumen2021}. 
From our literature analysis, general-purpose game engines such as Unity and Unreal are already used in approximately 60\% of relevant works on game engines for scientific visualization, with reported benefits in visual quality, interactivity, and communication of results~\cite{Greenwood2020,Fedotova2023}. In many of these studies, optimization methods originally developed for games (e.g., frustum culling, GPU instancing) are replicated or adapted to scientific settings~\cite{Fedotova2023}. Despite these advances, most Unreal-based systems are designed as bespoke applications focused on a single domain. Several domain-specific frameworks target immersive analysis in materials science and molecular visualization, emphasizing exploratory interaction and collaborative analysis, but they do not provide generic facilities for profiling and contrasting engine-level optimizations across multiple scenarios~\cite{Loeffler2019,Garcia2017,baaden2018unitymolapbs}. Recent surveys on immersive molecular visualization further highlight that many VR tools prioritize user experience, while performance tuning is handled ad hoc and seldom reported in a reproducible manner~\cite {baaden2023sotaVR,ruzic2016ivr}. Beyond game engines, the visualization community has investigated general frameworks for benchmarking interactive performance. Rossant et al.\ describe hardware-accelerated visualization pipelines with explicit GPU-centric optimizations, and Frey et al.\ propose a benchmarking framework for systematically evaluating the runtime performance of interactive visualizations under varying parameter configurations~\cite{rossant2013hardware,frey2019evaluating}. These works underline the importance of rigorous, experiment-based performance assessment, yet they mainly target custom OpenGL or in-house frameworks rather than mainstream game engines. UnrealVis is informed by these efforts but fundamentally differs in its methodological scope. While existing frameworks often hardcode specific optimization pipelines to maximize throughput for a target scientific domain (e.g., massive point clouds or fluid dynamics), they lack the infrastructure to systematically isolate, toggle, and quantify the performance impact of these techniques side by side. 
UnrealVis acts as a diagnostic game-engine-based laboratory, providing a generic, reproducible telemetry sandbox where rendering strategies, such as Nanite, explicit culling, and level streaming, can be dynamically combined and evaluated using a unified analytics pipeline. To the best of our knowledge, no existing system offers a similarly rigorous environment for empirically profiling optimization configurations across heterogeneous scientific datasets. Unlike domain-specific viewers, UnrealVis introduces a methodological shift: it treats optimization as a variable to be isolated and measured, rather than a hardcoded feature. This enables cross-domain performance comparisons that were previously unavailable in game-engine-based tools and helps assess their performance for scientific visualization.

\section{A Taxonomy of Optimizations}
\label{sec:req}

UnrealVis is grounded in a taxonomy of optimization techniques organized into six families, derived from a systematic literature review. This taxonomy bridges classical scientific visualization strategies with game engine capabilities, serving as both a design blueprint for the system and a framework for reproducible performance experiments.

\subsection{Method}
To derive the optimization taxonomy underlying UnrealVis, we conducted a focused literature study on rendering and simulation optimizations in scientific visualization and closely related domains. The goal was threefold: (i) identify optimization techniques that are explicitly used to address performance and scalability issues in scientific visualization, (ii) understand how these techniques can be instantiated in general-purpose game engines, and (iii) inform the design of a laboratory environment in Unreal Engine that can implement and compare such techniques systematically. We targeted works at the intersection of the following themes: \textit{scientific visualization}, \textit{game engines}, \textit{Unreal Engine}, \textit{molecular visualization}, \textit{materials science}, \textit{optimization techniques}, and \textit{performance}. Publications were retrieved from major scholarly indexes using combinations of these keywords and a time window spanning roughly 1990--2024. This process yielded 75 candidate papers; after title/abstract screening and full-text reading, 55 papers were analyzed in detail, and 20 were discarded as out of scope (e.g., purely theoretical turbulence modeling or visualization work without an explicit discussion of performance). For each selected paper, we extracted: (i) the optimization techniques explicitly mentioned (e.g., frustum culling, GPU ray casting, Nanite), (ii) the computational target of the technique, (iii) the primary goal (simulation acceleration versus interactive visualization), and (iv) the domain context~\cite{Ayachit2015,Loeffler2019,Ingo2016}. We then grouped techniques into families based on their primary computational focus and implementation characteristics, iteratively refining the grouping until we obtained six stable categories. From this survey, we observe that general-purpose engines such as Unity and Unreal Engine are already adopted in approximately 60\% of the selected papers that explicitly discuss game engines for scientific visualization~\cite{Greenwood2020}. In many of these works, optimization methods originally developed for games (e.g., LOD, culling, GPU instancing) are replicated or adapted to scientific settings~\cite{Fedotova2023}. Unreal Engine, in particular, offers a rich set of built-in optimizations and exposes APIs that facilitate implementing more advanced techniques~\cite{choi2024chimera}. The final taxonomy reflects both the frequency with which techniques appear in the literature and their practical feasibility in an Unreal-based laboratory.

\subsection{Taxonomy Structure}

\begin{figure}[h]
    \centering
    \resizebox{\columnwidth}{!}{
    \begin{tikzpicture}[
      basic/.style  = {draw=none, text width=1.2cm, font=\sffamily\tiny, rectangle, align=center},
      root/.style   = {basic, draw, font=\sffamily\bfseries\small, text width=7cm, fill=teal!80, minimum height=20pt},
      level-2/.style = {basic, draw, minimum height=28pt, text width=1.1cm, font=\sffamily\tiny\bfseries, fill=teal!50, anchor=north},
      level-3/.style = {basic, minimum height=25pt, text width=1.1cm, font=\sffamily\tiny, fill=teal!30, anchor=north},
      level 1/.style={sibling distance=4.3em, level distance=4em},
      edge from parent/.style={draw=none},
      node distance=1.1cm
      ]
      
    % Root
    \node[root] (rootNode) {Interactive Driven Optimization}
    child {node[level-2, fill={rgb,1:red,0.996;green,0.851;blue,0.412}] (c1) {Rendering \\ Optimization}}
    child {node[level-2, fill={rgb,1:red,0.431;green,0.906;blue,0.824}] (c2) {Shadow \\ Optimization}}
    child {node[level-2, fill={rgb,1:red,0.706;green,0.949;blue,0.733}] (c3) {Data \\ Optimization}}
    child {node[level-2, fill={rgb,1:red,0.839;green,0.710;blue,1.0}] (c4) {Geometry \\ Optimization}}
    child {node[level-2, fill={rgb,1:red,0.984;green,0.780;blue,0.714}] (c5) {CPU \\ Based}}
    child {node[level-2, fill={rgb,1:red,0.929;green,0.929;blue,0.929}] (c6) {Engine \\ Based}};

    % Level 3 Nodes
    \begin{scope}[every node/.style={level-3}]
    % Column 1
    \node [below=0.6cm of c1, fill=teal!40] (c11) {Ray Tracing};
    \node [below=0.1cm of c11, fill=teal!5] (c12) {GPU Shaders};
    \node [below=0.1cm of c12, opacity=0] (c13) {};
    \node [below=0.1cm of c13, fill=teal!20] (c14) {Volumetric \\ Rendering};
    \node [below=0.1cm of c14, fill=teal!5] (c15) {GPU \\ Instancing};
    \node [below=0.1cm of c15, fill=teal!5] (c16) {GPU \\ Raycasting};

    % Column 2
    \node [below=0.6cm of c2, fill=teal!20] (c21) {Ambient \\ Occlusion};
    \node [below=0.1cm of c21, opacity=0] (c22) {};
    \node [below=0.1cm of c22, opacity=0] (c23) {};
    \node [below=0.1cm of c23, fill=teal!20] (c24) {Virtual \\ Shadow};

    % Column 3
    \node [below=0.6cm of c3, fill=teal!20] (c31) {Data \\ Compression};
    \node [below=0.1cm of c31, fill=teal!20] (c32) {Instancing};

    % Column 4
    \node [below=0.6cm of c4, fill=teal!40] (c41) {LOD / HLOD};
    \node [below=0.1cm of c41, fill=teal!20] (c42) {Culling \\ Techniques};
    \node [below=0.1cm of c42, fill=teal!20] (c43) {Empty Space \\ Skipping};
    \node [below=0.1cm of c43, fill=teal!5] (c44) {Marching \\ Cubes};
    \node [below=0.1cm of c44, fill=teal!5] (c45) {Geometry \\ Clipmaps};
    \node [below=0.1cm of c45, fill=teal!20] (c46) {LOD};

    % Column 5
    \node [below=0.6cm of c5, fill=teal!5] (c51) {CPU Parallel};
    \node [below=0.1cm of c51, fill=teal!5] (c52) {ML Driven \\ Algorithm};

    % Column 6
    \node [below=0.6cm of c6, fill=teal!40] (c61) {Internal};
    \node [below=0.1cm of c61, fill=teal!20] (c62) {Nanite};
    \node [below=0.1cm of c62, fill=teal!5] (c63) {Level \\ Streaming};
    \node [below=0.1cm of c63, fill=teal!20] (c64) {Internal};
    \node [below=0.1cm of c64, fill=teal!20] (c65) {Lumen};
    \end{scope}

    % Vertical Grid Lines
    \foreach \x in {c1, c2, c3, c4, c5} {
        \draw[thick, draw=teal!30] ([yshift=0.4cm, xshift=0.68cm]\x.north) -- ([yshift=-8.8cm, xshift=0.68cm]\x.north);
    }

    % Category Labels
    \draw[thick, draw=teal!80] ([yshift=1.8cm, xshift=-0.1cm]c1.north west) -- ([yshift=1.8cm, xshift=0.1cm]c6.north east) 
    node[midway, above, font=\sffamily\tiny\bfseries, fill=white] {All Techniques};
    
    \draw[thick, draw=teal!50] ([yshift=0.4cm, xshift=-0.1cm]c1.north west) -- ([yshift=0.4cm, xshift=0.1cm]c6.north east) 
    node[midway, above, font=\sffamily\tiny\bfseries, fill=white] {Optimization Categories};

    % Specific Techniques Family Line
    \draw[thick, draw=teal!30] ([yshift=-0.25cm, xshift=-0.1cm]c1.south west) -- ([yshift=-0.25cm, xshift=0.1cm]c6.south east) 
    node[midway, fill=white, font=\sffamily\tiny\bfseries] {Specific Techniques Family};

    % Extra Vertical Layers
    \draw[thick, draw=teal!30] ([xshift=-0.35cm]c11.north west) -- ([xshift=-0.35cm]c13.south west)
    node[midway, font=\sffamily\tiny, fill=white, rotate=90] {Visualization};

    \draw[thick, draw=teal!30] ([xshift=-0.35cm]c14.north west) -- ([xshift=-0.35cm]c16.south west)
    node[midway, font=\sffamily\tiny, fill=white, rotate=90] {Simulation};

    % Legend
    \begin{scope}[xshift=4.5cm, yshift=-8.8cm]
        \node[anchor=west, font=\sffamily\tiny\bfseries] at (0,0.5) {Legend};
        \draw[thick, fill=teal!5] (0,0) rectangle (0.3,-0.2);
        \node[anchor=west,font=\sffamily\tiny] at (0.35,-0.1) {just one};
        \draw[thick, fill=teal!20] (0,-0.3) rectangle (0.3,-0.5);
        \node[anchor=west,font=\sffamily\tiny] at (0.35,-0.4) {at most 3};
        \draw[thick, fill=teal!40] (0,-0.6) rectangle (0.3,-0.8);
        \node[anchor=west,font=\sffamily\tiny] at (0.35,-0.7) {more than 4};
    \end{scope}

    \end{tikzpicture}
    }
    \vspace{-0.2cm}
    \caption{The UnrealVis optimization taxonomy: categories and specific techniques synthesized from the literature review and engine-native capabilities, divided into visualization-centric  (top) and simulation-centric methods (bottom). The color-coding assigned to the six primary families is mirrored in the UnrealVis user interface. The color legend reports how many times the optimization techniques appear in the literature.}
    \label{fig:taxonomy}
\end{figure}
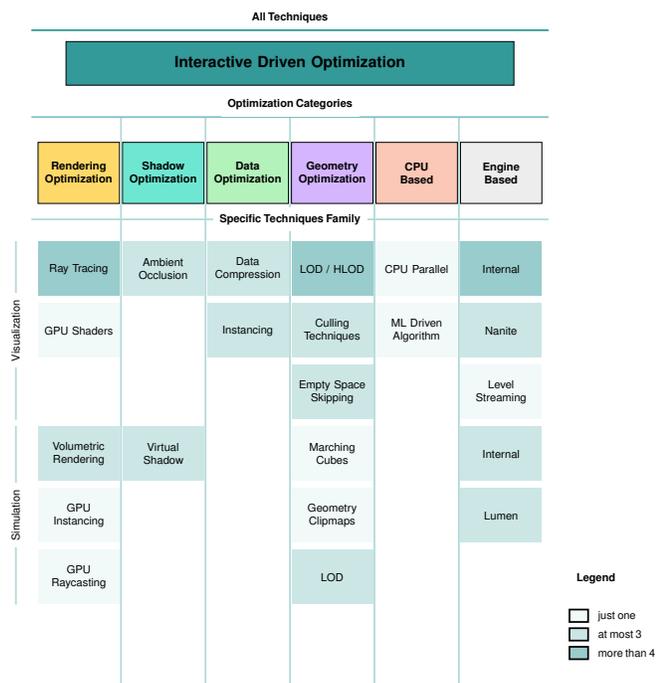

As illustrated in Figure~\ref{fig:taxonomy}, techniques are categorized along two operational dimensions: \textbf{Visualization-centric} methods (top), which primarily optimize the final pixel synthesis and shading, and \textbf{Simulation-centric} methods (bottom), which focus on the underlying data management and engine-level processing. To ensure cognitive continuity between this theoretical framework and our practical implementation, the six families are assigned a specific color-coding scheme. This exact palette is integrated into the UnrealVis graphical user interface (see Figure~\ref{fig:workflow}B for more details), allowing researchers to intuitively group, identify, filter, and toggle related optimizations during their experimental setups. The proposed taxonomy is detailed below:

\vspace{0.2cm}

\par\noindent\tikz[baseline=0.1em]\draw[black,fill={rgb,1:red,0.996;green,0.851;blue,0.412}] (0,0) rectangle (0.8em, 0.8em); \textbf{Rendering Optimization (5 techniques).} These strategies accelerate pixel-level calculations and shading by optimizing how light and materials are computed to ensure high-performance throughput in visually dense scenes \cite{reina2020mtv}. Beyond traditional \textit{ray tracing} \cite{Ingo2016}, we identify \textit{ray marching}, which traverses volumetric data or scalar fields step-by-step to render complex datasets \cite{Laureanti2020, kruger2003hardware}, and \textit{GPU raycasting} for the interactive representation of Solvent-Excluded Surfaces (SES) \cite{Martinez2019}. Specialized shaders such as \textit{HyperBalls} replace standard cylinders with hyperboloid primitives to improve the representation of molecular bonds \cite{Chavent2011, UnityMol2013}, while \textit{GPU instancing} significantly reduces draw call overhead (the command sent from the CPU to the GPU) by rendering multiple instances of identical geometry in a single hardware pass \cite{LeMuzic2015, Ferreira2018}.

\par\noindent\tikz[baseline=0.1em]\draw[black,fill={rgb,1:red,0.431;green,0.906;blue,0.824}] (0,0) rectangle (0.8em, 0.8em); \textbf{Shadow Optimization (2 techniques).} These methods approximate global illumination (GI) and depth cues while minimizing GPU costs. Shadows are critical for spatial perception, as they allow users to intuitively grasp the relative distance and scale of objects in 3D space \cite{reina2020mtv}. Techniques include \textit{Screen Space Ambient Occlusion} (SSAO), which calculates how much ambient light reaches a specific point to darken crevices and fine structural details \cite{Fedotova2023}, and \textit{Virtual Shadow Maps} (VSM), which provide high-resolution shadows across vast scales by dynamically allocating resolution based on the observer's current needs \cite{karis2021vsm}.

\par\noindent\tikz[baseline=0.1em]\draw[black,fill={rgb,1:red,0.706;green,0.949;blue,0.733}] (0,0) rectangle (0.8em, 0.8em); \textbf{Data Optimization (2 techniques).} This family focuses on reducing the overhead associated with data transfer and storage, a mandatory step when scientific datasets exceed available hardware memory limits \cite{Ayachit2015, reina2020mtv}. \textit{Data compression} enables real-time interaction with massive volumetric or time-series data by reducing their memory footprint \cite{Ayachit2015, wang2023lvdif}, while \textit{data instancing} optimizes performance by reusing repetitive structural elements, avoiding redundant memory allocations.

\par\noindent\tikz[baseline=0.1em]\draw[black,fill={rgb,1:red,0.839;green,0.710;blue,1.0}] (0,0) rectangle (0.8em, 0.8em); \textbf{Geometry Optimization (6 techniques).} These are core strategies for managing geometric complexity by lowering the number of triangles the GPU must process per frame. We distinguish between standard \textit{Level of Detail} (LOD) and \textit{Hierarchical LOD} (HLOD), which organizes the entire scene into spatial clusters to maintain interactive speeds in massive datasets \cite{garland2001multiresolution, ulrich2002hlod}. \textit{Visibility culling} ensures that the engine only processes objects that are actually visible to the observer \cite{chong2017gears}. Surface extraction methods like \textit{Marching Cubes} \cite{lorensen1987marching} and \textit{Empty Space Skipping} \cite{li2003emptyspace} further optimize volumetric traversal by ignoring regions that do not contain relevant data.

\par\noindent\tikz[baseline=0.1em]\draw[black,fill={rgb,1:red,0.984;green,0.780;blue,0.714}] (0,0) rectangle (0.8em, 0.8em); \textbf{CPU-Based Optimization (2 techniques).} These algorithms leverage general-purpose computation on the Central Processing Unit (CPU) to alleviate bottlenecks during data preprocessing and simulation management. Beyond \textit{parallel processing} on multi-core systems \cite{Ingo2016, rossant2013hardware}, \textit{Machine Learning} (ML) driven algorithms are increasingly used to refine computational processes, such as enabling interactive simulation steering once specific conformational states are reached \cite{lanrezac2021wielding}.

\par\noindent\tikz[baseline=0.1em]\draw[black,fill={rgb,1:red,0.929;green,0.929;blue,0.929}] (0,0) rectangle (0.8em, 0.8em); \textbf{Engine-Based Optimization (5 techniques).} These are native mechanisms provided by modern game engines like Unreal Engine 5 (UE5). \textit{Nanite} (virtualized geometry) enables pixel-scale detail by streaming only the detail perceptible to the camera \cite{EpicNanite2021, diazaleman2023nanite}. \textit{Lumen} provides dynamic global illumination that reacts instantly to scene changes \cite{EpicLumen2021}, while \textit{Level Streaming} manages the asynchronous loading and unloading of dataset sections to prevent memory overload in vast or complex environments.

\section{The UnrealVis System}
\label{sec:system}

The literature review and the resulting taxonomy highlight a critical gap: while many optimization techniques exist, they are typically hardcoded into domain-specific applications, making it difficult to systematically assess and compare their impact. To bridge this gap, we developed UnrealVis, a modular laboratory environment built natively in Unreal Engine 5.3.2 (UE5) to serve as a systematic testing ground for rendering optimization techniques in scientific visualization~\cite{reina2020mtv,Loeffler2019}. The architecture and interactive features of UnrealVis were driven by four primary design goals (DGs), formulated to address the challenges identified in our taxonomy:

\par\noindent\textbf{DG1: Interactive Exploration (Real-Time Performance).} The system must support smooth, real-time navigation of massive datasets by exposing and dynamically combining the engine-level geometry and visibility optimizations identified in the taxonomy.
%(i.e., Nanite, explicit culling, LOD).

\par\noindent\textbf{DG2: Scientific Accuracy (Visual Fidelity).} Performance gains must not compromise the interpretability of the data. The system must allow researchers to prioritize high-fidelity rendering in specific regions of interest (i.e., active molecular sites) while optimizing peripheral context.

\par\noindent\textbf{DG3: Reproducible Analytics (Performance Telemetry).} To serve as a true diagnostic laboratory, the system must decouple data exploration from performance profiling. It provides a frame-accurate sampling engine that generates standardized, exportable logs, enabling researchers to transform subjective exploration into a rigorous, quantitative benchmark of optimization profiles.

\par\noindent\textbf{DG4: Architectural Scalability (Generality).} The framework should be domain-agnostic, providing flexible data ingestion pipelines capable of handling diverse scientific formats, from atomistic point clouds to dense volumetric fluid dynamics.

Guided by these goals, the system is engineered to bridge high-fidelity, real-time rendering and the rigorous data-integrity requirements of scientific research~\cite{Ayachit2015}. To minimize computational overhead and avoid skewing performance telemetry, UnrealVis uses custom C++ modules and the Slate UI framework rather than high-level, blueprint-based abstractions such as UMG (Unreal Motion Graphics). The system logic centers on an asynchronous data pipeline for non-blocking ingestion of large-scale scientific datasets, specifically biological structures in PDB format~\cite{VMD1996}. A frame-accurate sampling engine monitors hardware metrics, including kernel-level CPU load, render-thread GPU timing, and physical RAM consumption, at intervals synchronized with the system clock. This telemetry allows researchers to isolate the impact of specific optimizations, such as culling or virtualized geometry, while a Comparative Analytics Module enables side-by-side quantitative validation through integrated charts. The interface facilitates exploration via a Full View Mode, which removes HUD elements to enable unhindered navigation of 3D structures along the \texttt{XYZ} spatial axes. Researchers can transition between qualitative data exploration and quantitative analysis, leveraging integrated charting tools to visualize trends and performance trade-offs directly within the simulation environment~\cite{frey2019evaluating}.

\subsection{Implementation}
\label{sec:implementation}

The implementation of UnrealVis addresses common challenges like visual overload and limited scalability by integrating native UE5 features~\cite{marsden2020unreal}. Culling techniques (Frustum and Occlusion) are utilized to exclude non-visible elements from the pipeline~\cite{chong2017gears}, while LOD management dynamically adjusts geometric complexity. Advanced systems like Nanite provide virtualized geometry that automatically handles polygon density at a micro-level, ensuring high visual fidelity without the manual overhead of mesh decimation~\cite{EpicNanite2021}.

\subsubsection{Data Ingestion and Multiphysics Transcoding}

\begin{figure}[htbp]
    \vspace{-2mm}
    \centering
    \includegraphics[width=\columnwidth]{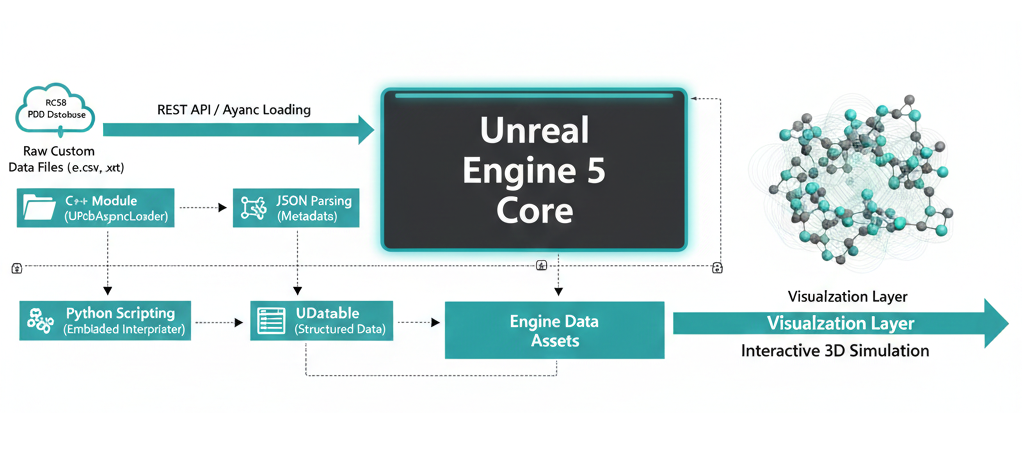}
    \vspace{-2mm}
    \caption{Technical schematic of the UnrealVis data ingestion pipeline, highlighting the dual-path architecture: the asynchronous C++ module for PDB retrieval and the Python-integrated scripting layer for custom dataset processing and OpenEXR multiphysics transcoding.}
    \label{fig:ingestion_pipeline}
\end{figure}

UnrealVis employs a dual-path data ingestion pipeline (visible in Figure~\ref{fig:ingestion_pipeline}) designed to balance flexibility with high-performance GPU throughput. The primary path handles PDB structures through a native asynchronous C++ module. For complex simulation outputs, such as the BLASTNet $Re=5000$ lifted-hydrogen jet flame, the system uses a specialized \textit{Multiphysics Data Transcoding} pipeline. To overcome the I/O latency and memory bottlenecks inherent in raw ASCII buffers (e.g., \texttt{.dat} files), we transcode discrete physics variables into 32-bit Floating Point OpenEXR (EXR) textures \cite{blastnet2023neurips}. This allows the engine to treat physical states as spatial signal data. We map four critical properties to the high-dynamic-range (HDR) color channels: horizontal momentum ($U_x$) to Red, axial thrust ($U_y$) to Green, temperature ($T_K$) to Blue, and chemical reactivity (OH mass fraction) to Alpha. This 1:1 mapping ensures that the GPU can sample a complete physical state at any spatial coordinate in a single hardware-native pass, preserving numerical precision while significantly reducing storage footprint and CPU overhead.

\subsection{Domain-Specific Rendering Techniques}
\label{sec:visual_representation}

The visualization layer of UnrealVis converts abstract coordinate data into intuitive 3D representations optimized for both visual clarity and rendering performance. Each dataset type follows a tailored representation strategy that balances scientific accuracy with real-time interactivity. 
For molecular structures from the PDB, atoms are rendered as GPU-instanced static meshes: spheres represent atomic positions while thin cylinders depict molecular bonds in ribbon-style secondary structure visualization~\cite{VMD1996}. Large complexes such as the bacterial 70S ribosome (8B0X, $\sim$500k atoms) or eukaryotic 80S ribosome (4V6W, $\sim$1M atoms)~\cite{pdb_8b0x,pdb_4v6w} are automatically partitioned into hierarchical LOD groups during ingestion. Core functional regions (active sites, binding pockets) receive high-detail representations even when zoomed out, while peripheral structural elements use progressively simplified geometry. This spatial LOD assignment ensures that researchers can maintain focus on biologically relevant regions without the GPU being overwhelmed by distant structural context.

For volumetric fluid dynamics and turbulent flow fields sourced from the BLASTNet 2.0 repository~\cite{blastnet2023neurips}, the visualization strategy utilizes the \textit{Multiphysics Data Transcoding} pipeline. Data is fed into GPU-driven Niagara systems, which sample the transcoded HDR textures to animate particles along the exact streamlines of the original simulation. This eliminates the need for simulated forces, as particles follow the pre-calculated velocity vectors ($U_x, U_y$) stored in the texture channels. Temporal dynamics are handled by importing individual snapshots as a \textit{Texture 2D Array}, allowing the engine to interpolate between ``Z-slices'' at runtime for a fluid, time-accurate representation of turbulence flickering. Domain experts can visualize complex flow features, such as the ``lift-off'' height in hydrogen flames, by mapping physical scalars (e.g., $T_K$, OH fraction) directly to emissive color curves within the viewport.

For immersive exploration, UnrealVis provides Full View Mode, which strips away all HUD elements to create an uncluttered viewport optimized for spatial navigation. Standard \texttt{XYZ} camera controls (\texttt{WASD} translation, mouse-look rotation) enable researchers to ``fly through'' dense molecular assemblies or volumetric fields, with performance telemetry remaining active in a minimal overlay to support data-driven navigation decisions.

\subsection{Optimization Techniques}

UnrealVis is designed as a small laboratory for testing how different rendering optimizations affect the interactive exploration of complex 3D scientific datasets. Building on the taxonomy of techniques discussed in the literature \cite{diazaleman2023nanite,reina2020mtv}, the system focuses on a subset that can be combined and toggled at runtime to study their impact on both performance and visual quality. In particular, UnrealVis integrates geometry simplification, visibility reduction, and memory–aware scene management in a coherent workflow, rather than exposing them as isolated engine features. From a geometric perspective, UnrealVis relies on Unreal Engine 5’s Nanite virtualized geometry system whenever possible, allowing highly detailed meshes to be rendered by streaming only the detail that is perceptible to the camera \cite{EpicNanite2021}. For assets that cannot use Nanite, the framework provides a custom LOD strategy tailored to scientific data: different portions of the structure are assigned to separate LOD levels according to their importance and typical viewing distance. This enables large molecular assemblies or dense point clouds to retain full detail in the focus region, while distant elements are represented with progressively simpler geometry, reducing the number of processed triangles without compromising the overall readability of the visualization. To further limit the amount of work sent to the GPU, UnrealVis combines several visibility-culling strategies. Native frustum and occlusion culling are exploited to discard objects that lie outside the camera view or are fully hidden behind other geometry \cite{chong2017gears}. On top of these engine-level mechanisms, UnrealVis introduces an explicit distance–culling layer: each actor can be assigned a maximum draw distance, so that very distant atoms, particles, or auxiliary meshes are never submitted to the renderer. This is particularly effective in scenes with millions of small repeated elements, where many objects contribute little to the final image but impose a high cost in terms of draw calls and overdraw. Finally, UnrealVis uses Level Streaming to manage memory usage when visualizing very large environments or multi-part datasets. Instead of loading the entire scene at once, data are partitioned into logical sub-levels (for example, different regions of a ribosome or disjoint sub-volumes of a turbulence field). These sub-levels are loaded and unloaded dynamically as the user moves through the scene, keeping the working set of geometry and textures within the limits of commodity hardware while still supporting seamless exploration.

\subsection{Telemetry and Sampling Metrics}

To evaluate these optimization techniques in a reproducible way, UnrealVis includes a high-frequency sampling engine that monitors a compact set of performance indicators during each simulation session. Frame-based metrics are derived directly from engine timing: the system computes FPS as the inverse of the per-frame \texttt{DeltaTime} and stores the raw frame time in milliseconds, enabling detection of both sustained slowdowns and short micro-stutters. In parallel, UnrealVis queries the operating system to obtain CPU utilization (using PDH counters on Windows) and reads memory statistics via \texttt{FPlatformMemory}, recording the amount of physical RAM used by the process. On the graphics side, render-thread timing information (e.g., \texttt{GRenderThreadTime}) is sampled to approximate GPU frame time and to separate rendering load from general game-thread activity. All sampled metrics are written to disk in JSON format through the \texttt{USimulationExporter} module. Each simulation session thus produces a self-contained log that can be reloaded inside UnrealVis for chart-based comparison or inspected externally with standard data analysis tools. This design mirrors best practices in existing profiling workflows \cite{Fedotova2023,choi2024chimera}, but packages them in a form that is directly accessible to domain experts who may not be familiar with low-level engine profilers.

\subsection{Comparative Analytics Algorithm}
\begin{figure*}[t!]
    \vspace{-2mm}
    \centering
    \includegraphics[width=\textwidth]{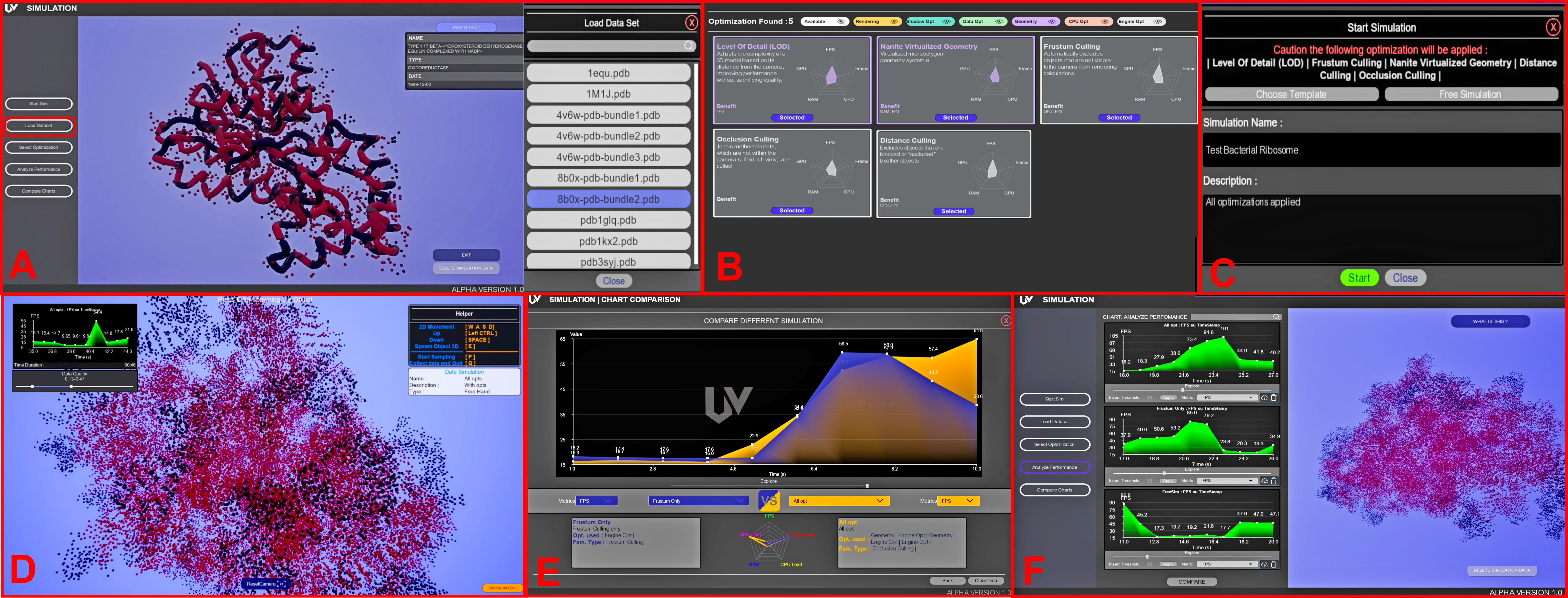}
    \vspace{-2mm}
    \caption{An example workflow for starting a simulation in UnrealVis, showing a huge bacterial 70S ribosome. The user is first greeted by a welcome screen, displaying a human 17$\beta$-hydroxysteroid dehydrogenase type 1 (oxido-reductase enzyme, $\sim$35 kDa) complexed with NADP$^+$ cofactor and equilin inhibitor \cite{pdb_1equ, Sawicki1999} by default. A menu shows the different views selection on the left, while a small panel on the right briefly describes the loaded dataset. To begin the workflow, the user selects the dataset from the list (A), then chooses the desired optimizations from the appropriate menu (B). After renaming the simulation and inputting its description (C), the simulation can start. The user can freely move within the 3D dataset using the mouse and keyboard, while performance metrics are captured in real time (D). Then, the user can analyze performance statistics relative to simulations by directly comparing two simulations over time (E) or by inspecting performance using small multiples when simulations are many (F).}
    \label{fig:workflow}
\end{figure*}
On top of raw telemetry, UnrealVis implements a lightweight comparative analytics layer that powers the ``VS'' (comparison) views shown in the chart interfaces. For every completed simulation, the \texttt{USimulationAnalytics} module computes the arithmetic mean of each metric over the session duration, enabling fair comparisons even when simulations differ in length or camera trajectory. These normalized statistics are then used to define simple, metric-dependent victory conditions: for FPS, higher mean values indicate better performance, whereas for CPU load, RAM usage, and frame time, lower values are preferable. By applying these rules, the system can automatically determine which optimization configuration ``wins'' on each metric. The results are presented through synchronized line charts and summary indicators (implemented with own written plugin and the HUD), allowing users to see both the temporal evolution of metrics and the aggregate comparison between simulations. When combined with the two-way whisker, which isolates specific spatial or temporal intervals of interest, this comparative view helps distinguish optimizations that are globally effective across an entire session from those that primarily improve local behavior in a subset of camera positions or data densities. In practice, UnrealVis supports a workflow in which researchers can iteratively test different combinations of Nanite, LOD, culling, and streaming and immediately see how each choice affects performance and visual quality at both global and local scales.

\subsection{Supported Datasets}
The initial version of UnrealVis focuses on two representative dataset families: atomistic biomolecular structures from the PDB~\cite{Berman2000PDB} and dense 3D fields derived from turbulent combustion simulations. Together, these datasets span several orders of magnitude in geometric complexity, from tens of thousands to several hundred thousand primitives per frame, and are used throughout our case studies to exercise different optimization configurations. For biomolecular data, UnrealVis natively supports a curated set of PDB entries chosen to cover a range of sizes and structural organizations. At the high end of complexity, we include the bacterial 70S ribosome (8B0X), a translating \textit{Escherichia coli} ribosome in the unrotated state with P- and E-site tRNAs ($\sim$2.5~MDa, $\sim$500k atoms), which serves as our ``stress-test'' dataset for large macromolecular assemblies~\cite{pdb_8b0x,Fromm2023}. As a metazoan counterpart, we use the \textit{Drosophila melanogaster} 80S ribosome (4V6W), available in multiple preprocessed ``bundles'' that differ in the number of included chains, providing medium- to large-scale test cases while keeping the molecular architecture comparable across runs~\cite{pdb_4v6w}. Smaller, enzyme-scale structures are represented by the human 17$\beta$-hydroxysteroid dehydrogenase type~1 in complex with NADP$^+$ and equilin (1EQU, $\sim$35~kDa) and by chymotrypsin inhibitor complexes (e.g., 1GLQ-like structures), which offer smooth navigation scenarios suitable for assessing the overhead of individual optimizations at lower geometric density~\cite{pdb_1equ,Sawicki1999}. Finally, medium-sized adhesin structures such as the \textit{Haemophilus influenzae} Hap adhesin (3SYJ) provide elongated, filamentous geometries that are particularly informative when evaluating culling and distance-based optimizations during close-up fly-throughs~\cite{pdb_3syj,Meng2011}. Beyond molecular systems, UnrealVis is being extended to support turbulent flow datasets inspired by the BLASTNet benchmark. In particular, we preprocess sub-volumes from hydrogen–air DNS studies into point clouds and volumetric fields, mapping grid points to particles and encoding scalar quantities such as mixture fraction, progress variable, or vorticity magnitude into color and size attributes~\cite{blastnet2023neurips}. These derived fields are representative of dense volumetric flow data encountered in the BLASTNet “diluted partially premixed \mbox{H\textsubscript{2}/air} lifted flame” configurations, and they are used to test optimization strategies that combine particle LOD, frustum and distance culling, and level streaming under highly cluttered visual conditions. Across these datasets, our goal is not to exhaustively cover all possible scientific domains, but rather to provide a controlled yet diverse set of benchmarks in which UnrealVis can apply its optimizations and comparative analytics to both molecular and flow-centered scenarios.

\subsection{Experimental Workflow}
\label{sec:workflow}

UnrealVis allows starting a simulation with a chosen dataset to analyze and explore it in depth by trying various optimization combinations and seeing the exploration performance in real time. A complete workflow example can be seen in Figure~\ref{fig:workflow}.

\par\noindent\textbf{Simulation Setup} 
The user first loads the dataset from a menu (Figure~\ref{fig:workflow} A). In this case, the selected dataset is 8B0X bundle 2, consisting of a massive bacterial 70S ribosome ($\sim$2.5 MDa, $\sim$500k atoms) in its translating, unrotated state, rendered as a dense 3D point cloud for performance testing \cite{pdb_8b0x,Fromm2023}. Then, the user selects the desired optimizations to test from the dedicated menu, which comprises rendering, shadow, data, geometry, CPU, and engine optimizations (Figure~\ref{fig:workflow}B). Each optimization comes with a description, expected performance benefits, and a radar chart that provides quick insight into its impact across FPS, frame time, CPU, RAM, and GPU usage. Optimizations can be filtered by type and availability directly within UnrealVis. After these steps, pressing the \texttt{Start Sim} button opens a panel that shows the applied optimizations and allows the user to save the simulation name and description for further analysis (Figure~\ref{fig:workflow} C). It is also possible to choose a preset template consisting of different optimization combinations, or simply navigate through the 3D dataset in a free simulation scenario.

\par\noindent\textbf{Real Time exploration and telemetry} 
After selecting the desired optimizations, the user can start the simulation. Pressing \texttt{E} spawns the dataset, which can then be navigated using the controls shown in the right panel (Figure~\ref{fig:workflow} D). The \texttt{W A S D} keys enable 2D camera movement, while mouse drag gestures rotate the view to inspect the dataset from any desired perspective. To begin performance data collection, the user presses \texttt{P}, which activates real-time sampling and captures FPS, frame time, CPU load, RAM usage, and GPU frame time. UnrealVis displays a dynamic line chart in the top-left corner that tracks FPS evolution over time, complemented by a 2-way whisker to focus on specific intervals in the dataset. This allows balancing data representation quality against rendering performance to identify optimal settings for fluid dataset exploration. Pressing the \texttt{Q} key will stop the simulation and save the results, ready for inspection.

\subsubsection{Standardized Simulation Templates}
\label{sec:templates}

To ensure reproducible performance profiling and mitigate the inherent variability of manual human exploration (\textbf{DG3}), UnrealVis supports the use of standardized simulation templates (selectable from the optimization menu, Figure \ref{fig:workflow} C), employing fixed camera paths. By constraining the camera to predefined paths, the system allows researchers to isolate the rendering cost of specific optimization configurations under identical workloads. The framework currently provides three primary templates, each designed to stress-test different components of the rendering pipeline, as described later in Section (\ref{sec:benchmarks}):

\par\textbf{T1: Spawn (30 seconds):} This template focuses exclusively on the initial dataset loading and object instantiation. By maintaining a static camera position immediately after spawning, it provides a controlled baseline for evaluating the effectiveness of LOD and Nanite virtualization. This duration allows the system to stabilize and record the ``cost of readiness'' for the geometry without the interference of dynamic visibility changes.
\par\textbf{T2: LookAround (30 seconds):} This template executes a deterministic orbital rotation around the dataset. The movement is specifically designed to trigger frequent updates in the engine's visibility set, forcing the application of various culling levels, including Frustum and Occlusion Culling. It serves to identify the ``threshold of utility'' at which the CPU overhead of visibility calculation equals the actual GPU rendering gains.
\par\textbf{T3: Stress Test / Full Cycle (180 seconds):} This comprehensive template combines multiple phases to simulate a complete research workflow. It begins with a 60-second spawn phase to evaluate initial geometry simplification. This is followed by a 40-second external rotation to assess global structural representation. At the 100-second mark, the camera performs an internal fly-through for approximately 80 seconds, penetrating the dense core of the assembly to maximize pixel overdraw and test culling performance in highly cluttered conditions. The cycle concludes with a return to the final position, at which point the simulation results, comprising FPS, frame time, CPU/GPU load, and RAM usage, are collected and exported as JSON.

These templates complement the exploration process, enabling users to choose between unconstrained manual simulations and expedited, automated workflows.

\subsubsection{Analyzing Simulation Results}

Results can be analyzed in two ways: by opening the \texttt{Compare Charts} menu for a more fine-grained comparison of two simulations, or, if the simulations are many, by opening the \texttt{Analyze Performance} Menu for gaining quick insights. The first option allows a direct comparison between up to two simulations, selecting the desired metric (e.g., FPS), and inspecting the charts over time. For example, in Figure~\ref{fig:workflow} E, two simulations of the same dataset (8B0X bundle 2) are compared, showing that the right-most simulation, leveraging all the available optimizations in UnrealVis, performs slightly better than applying only the frustum culling optimization (left-most simulation). The second option allows for quick comparison of multiple simulations at the same time, displaying small-multiple visualizations that can be explored over time using the appropriate slider. In Figure~\ref{fig:workflow} F, three simulations of the 8B0X bundle 2 dataset are compared: the two just analyzed in the \texttt{Compare Charts} menu, and another one, corresponding to the free simulation option, which does not use any optimization. Indeed, choosing no optimization results in a significant drop in FPS. It is also possible to insert a threshold (e.g., 30 FPS) into the small multiples to see how the simulations perform relative to it.
Finally, UnrealVis allows downloading simulation data in JSON format, comprising metadata (name, description), whether a template was used, the set of optimizations applied, and the five metrics captured by UnrealVis: FPS, frame time, CPU load, RAM usage, and GPU frame time. This allows users to analyze their simulation performance with fine-grained precision.

\section{Evaluation}
\label{sec:eval}

To demonstrate the efficacy of UnrealVis as a diagnostic laboratory, we present a multi-tiered evaluation. First, we describe two case studies illustrating how the system supports domain-specific exploration workflows. Second, we report the findings from a formative expert evaluation assessing the tool's overall usability and efficacy. Finally, to address the variability inherent in human interaction and provide rigorous quantitative metrics, we introduce an automated benchmarking suite that uses fixed camera paths and leverages the templates described in Section \ref{sec:templates}.

\par\noindent\textbf{Formative Expert Evaluation Protocol} For the qualitative evaluation, we adopted a \textit{Think Aloud} protocol \cite{Ericsson1993}. Participants (N=4) were asked to verbalize their reasoning and frustrations as they navigated the datasets. This allowed us to correlate specific UI friction points with the observed SUS scores \cite{brooke1996sus}, identifying ``cognitive bottlenecks'' in the manual tuning of low-level engine parameters.

\subsection{Case Studies}
\label{sec:casestudies}

To demonstrate the efficacy of UnrealVis as a diagnostic laboratory, we present two case studies informed by real-world scientific workflows. These studies illustrate how domain experts can leverage the system to navigate the complex trade-offs between rendering performance and scientific data integrity, fulfilling the requirement for domain-agnostic scalability (DG4).

\subsubsection{Case Study 1: Adhesin Exploration}

\begin{figure}[h]
    \vspace{-2mm}
    \centering
    \includegraphics[width=\columnwidth]{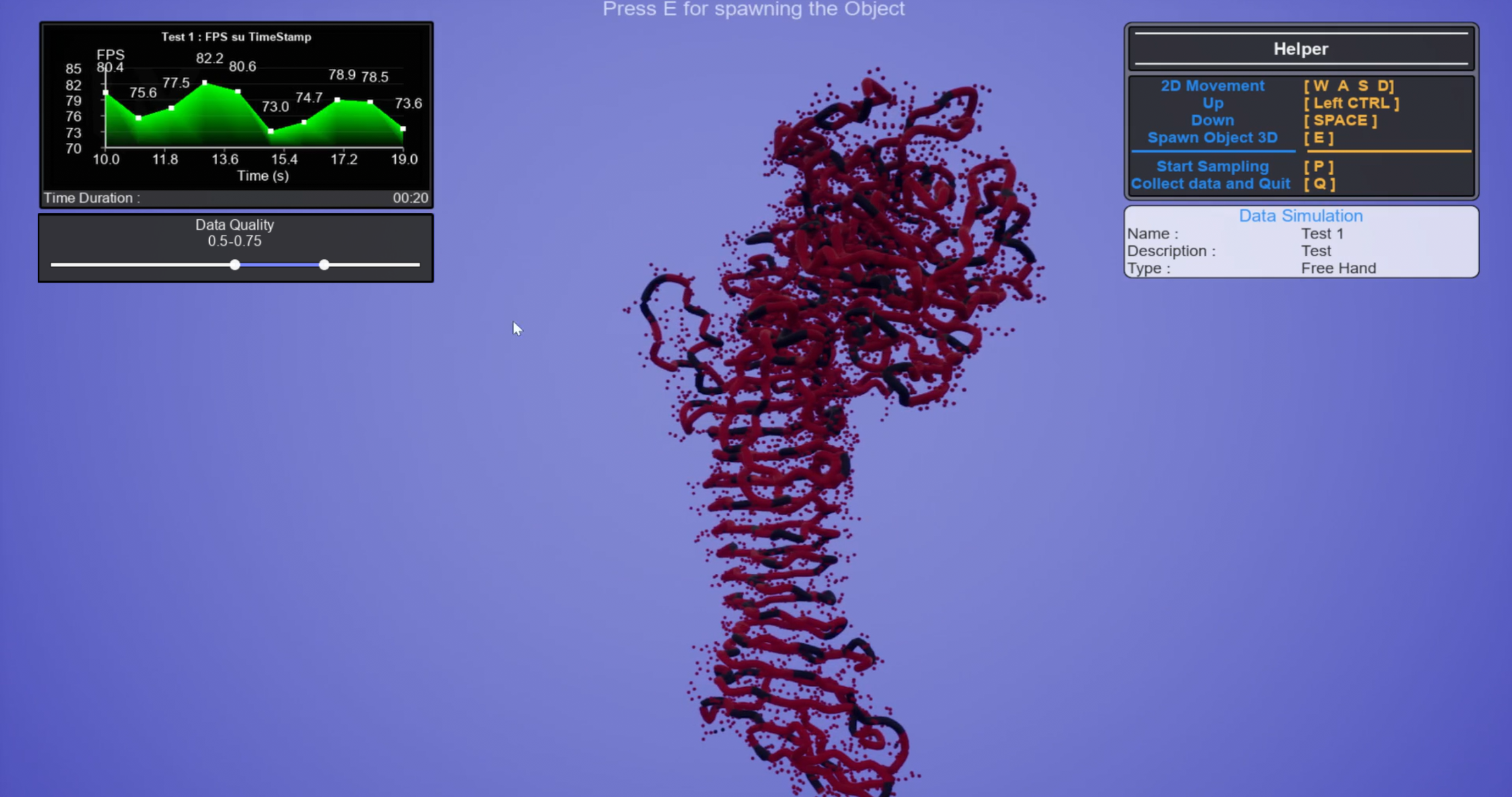}
    \vspace{-2mm}
    \caption{3SYJ adhesin exploration. The two-way whisker isolates the oligomerization interface (black), balancing high-fidelity rendering with global navigation performance.}
    \label{fig:casestudy_adhesin}
\end{figure}

A structural biologist investigates the oligomerization interface of the Haemophilus influenzae Hap adhesin (3SYJ\cite{pdb_3syj}, visible in Figure \ref{fig:casestudy_adhesin}).
In this scenario, the researcher is specifically interested in analyzing the protein's external surface and distal domains to identify potential interaction sites with the extracellular matrix. The primary challenge lies in the protein's elongated ``ribbon-and-sphere'' structure; while the researcher needs 100\% geometric fidelity on these outer regions, the sheer density of the underlying molecular core often cripples the fluidity of navigation. Following the UnrealVis workflow (Figure \ref{fig:workflow}), the biologist first loads the dataset (A), starts a free simulation (C), and conducts a diagnostic baseline simulation (D). At first glance, the live telemetry panel shows relatively stable average frame rates, which might suggest a well-performing environment; however, the scientist notices a persistent stutter during rapid orbital movements around the central domain that the aggregate FPS counter fails to fully explain. To investigate this discrepancy, the researcher leverages UnrealVis’s ability to export simulation data in JSON format for fine-grained precision analysis (supporting DG3) via the Analyze Performance menu (Figure \ref{fig:workflow}F). By inspecting the exported logs, which decouple game-thread activity from render-thread timing, the scientist identifies the true culprit: massive GPU frame-time spikes exceeding 1100 ms occurring during visibility updates. This micro-stuttering, objectively confirmed by the system's high-frequency sampling engine, renders smooth, intuitive exploration impossible. Leveraging this data-driven insight, the researcher proceeds to the optimization menu (Figure \ref{fig:workflow} B) to apply a combined strategy of Frustum Culling and LOD optimizations, guided by the menu's descriptions. While exploring and moving in the new simulation, the researcher notices that sliding one of the two whiskers isolates certain areas of the dataset, coloring them black and reducing their detail. Most importantly, the scientist discovers that the two-way whisker tool enables a ``precision strike'' on performance: the researcher isolates the central domain in black while rendering the periphery in red. By saving this simulation and comparing it to the baseline (E), the biologist confirms a 75\% reduction in maximum GPU spikes (from 1319 ms to 329 ms), achieving a stable 78 FPS without stuttering. The combination of the optimizations and the two-way whisker tool allowed the scientist to focus only on specific areas of interest in the dataset, keeping the exploration smooth (which validates DG1) and preserving the cognitive flow of the exploration, without sacrificing quality in the areas of interest (fulfilling DG2).

\subsubsection{Case Study 2: BLASTNet Exploration}

\begin{figure}[h]
    \vspace{-2mm}
    \centering
    \includegraphics[width=\columnwidth]{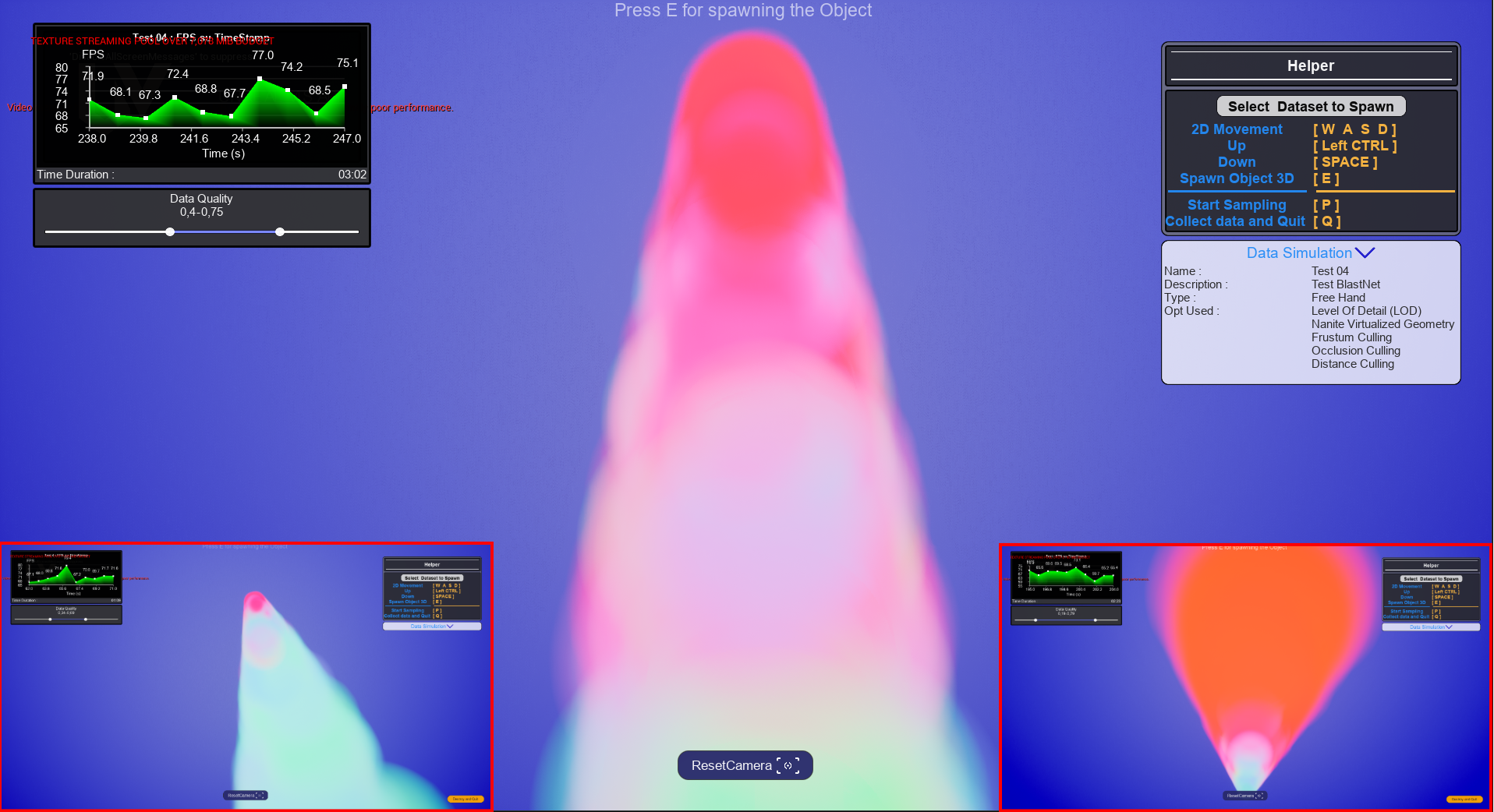}
    \vspace{-2mm}
    \caption{BLASTNet 2.0 volumetric exploration. The main view tracks real-time telemetry and optimization settings. Bottom insets provide multi-angle perspectives: the left inset highlights internal structures using a cyan-to-magenta gradient, while the right inset captures the high-vorticity red core from a wide-angle frontal view.}
    \label{fig:casestudy_hydrogenflames}
\end{figure}

The second case study focuses on engineers analyzing BLASTNet 2.0 ``Canonical Decaying HIT'' snapshots (Figure \ref{fig:casestudy_hydrogenflames}). %$\sim$1M particles
In this scenario, the objective is to trace coherent vortex structures within cluttered volumetric fields, specifically focusing on cells above the 95th vorticity percentile to study turbulence decay (supporting DG4). The primary challenge arises from the dense particle cloud, which, in a standard viewer, causes erratic frame pacing and prevents the engineer from following the temporal evolution of eddies. The engineer loads the dataset (\ref{fig:workflow}A) and starts an initial baseline navigation (D). While exploring the dataset with 3D movement, the engineer reveals a performance-stability conflict: although the average FPS appears functional, the dense volumetric traversal results in significant overdraw. To investigate the hardware impact without leaving the environment, the engineer accesses the Compare Charts menu (Figure \ref{fig:workflow} E). By overlaying the baseline telemetry against a reference performance target, the synchronized line charts reveal that the GPU fill rate is struggling with massive pixel overdraw, indicating a ``threshold of utility'' at which raw rendering power is overwhelmed by the unoptimized cloud. Leveraging these insights, the engineer accesses the optimization menu (B) to implement a multi-layered profile including Geometry Virtualization (Nanite) and Distance Culling. During the interactive session, the scientist uses the two-way whisker tool to perform a targeted local optimization: the high-vorticity cores are isolated in black (while maintaining full particle density and physical precision), while the surrounding ambient volume is heavily culled. The \textit{Analyze Performance} view (F) becomes crucial for identifying a fundamental hardware trade-off: while these optimizations successfully cap the maximum frame time at 72 ms (eliminating the 1.2-second freezes observed in the baseline), they increase peak RAM usage by approximately 200 MB to accommodate hierarchical data structures. The system enables identification of an optimal performance ceiling at which vortex topology is preserved despite aggressive culling of the surrounding volume, ensuring that scientific context is never compromised (fulfilling DG2). The final profile is then exported as a reusable template, transforming an ad-hoc tuning process into a reproducible, data-driven visualization strategy that maintains fluid navigation (validating DG1) even during internal fly-throughs.

\subsection{Formative Expert Evaluation}
\label{sec:usability}

To qualitatively assess UnrealVis's usability and identify areas for improvement, we conducted a formative evaluation with four users (Ev1, Ev2, Ev3, and Ev4) who have prior experience with both 3D game engines and scientific visualization. The sessions lasted approximately 90 minutes. Participants completed a short tutorial before performing six exploratory tasks covering the main UnrealVis functionalities. Feedback was gathered continuously via a Think-Aloud protocol~\cite{Ericsson1993}, supplemented by post-task questionnaires (Figure~\ref{fig:likert}) and a System Usability Scale (SUS) ~\cite{brooke1996sus}. We report the tasks below.

\par\noindent\textbf{Task 1 -- Starting the first free simulation.} Participants loaded the small 1EQU dataset~\cite{pdb_1equ}, inspected its description, and started a free simulation without optimizations, spawning the 3D structure and freely navigating it while recording FPS and other performance metrics.
\par\noindent\textbf{Task 2 -- Starting a complex free simulation.}
Users repeated the same procedure with the more demanding 8B0X bundle-2 dataset~\cite{pdb_8b0x}, again without optimizations, to experience baseline performance when exploring a very dense molecular assembly.

\par\noindent\textbf{Task 3 -- Starting an optimized simulation.}
Keeping the 8B0X dataset, evaluators opened the optimization menu, examined descriptions and radar charts for the available techniques, enabled Frustum Culling and LOD, and then ran a new simulation to observe how these optimizations affected navigation and performance. The evaluators used the whisker tool to mark some regions of the dataset in black. This allowed them to focus on specific portions of the dataset.

\par\noindent\textbf{Task 4 -- Performance comparison.}
Participants used the \emph{Compare Charts} view to load the two simulations from Tasks~2 and~3, scrolled the temporal slider, and identified intervals where one configuration outperformed the other in terms of FPS and frame time.

\par\noindent\textbf{Task 5 -- Testing a custom optimized simulation.}
Users configured a personalized combination of optimizations for the 8B0X dataset, based on their understanding of the radar charts and textual descriptions, and launched an additional simulation to assess whether their chosen settings improved or worsened interactive performance.

\par\noindent\textbf{Task 6 -- Comparing multiple simulations.}
Finally, evaluators opened the \emph{Analyze Performance} view to inspect all simulations of the 8B0X dataset at once, explored their evolution over time using a slider, applied a FPS threshold (e.g., 30 FPS), and saved selected simulation statistics for later inspection.

\subsubsection{Expert Feedback and Observations}
\label{sec:results}

\begin{figure}[h]
    \vspace{-2mm}
    \centering
    \includegraphics[width=0.8\columnwidth]{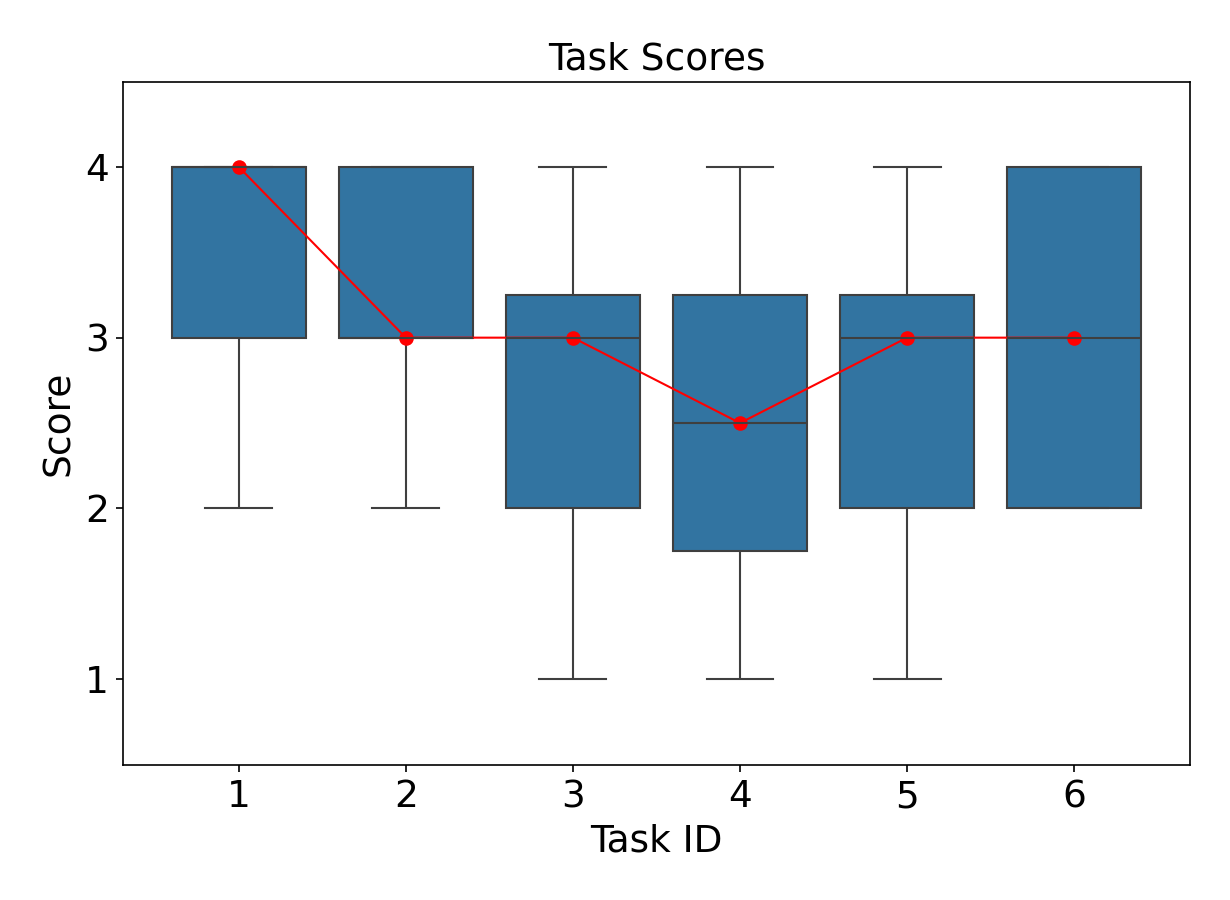}
    \vspace{-2mm}
    \caption{Aggregated Likert scale responses ($N=4$). High scores in Tasks 1--2 confirm the system's ease of use for general navigation. The score decrease in Tasks 4--5 (analysis and comparison) highlights the limitations of human perception in quantifying optimization impacts during manual interaction, justifying the need for deterministic benchmarks.}
    \label{fig:likert}
\end{figure}

The qualitative evaluation provided significant insights into the system's effectiveness and the inherent challenges of performance profiling across the six assigned tasks. Starting with the baseline exploration of the small 1EQU dataset \textbf{(Task 1)}, all evaluators ($N=4$) reported high ease of use (mean Likert > 3.5/4), quickly familiarizing themselves with the navigation and the live performance panel. However, the complexity increased significantly with the 8B0X dataset \textbf{(Task 2)}; while most users perceived a general lag, Ev4, leveraging a ``Very High'' expertise in 3D engines, specifically identified micro-stuttering during rapid camera rotations, noting that ``I noticed severe stuttering and impressive frame drops during complex movements, which makes smooth orbit and zoom impossible without optimizations.'' This was objectively confirmed by telemetry showing GPU frame-time spikes exceeding 1100 ms and severe frame drops to 2.5 FPS. When transitioning to the optimized simulation \textbf{(Task 3)}, the application of Frustum Culling and LOD schemes yielded a dramatic improvement in stability; the simulation logs showed a reduction in peak GPU frame times from approximately 1319 ms in the baseline to 329 ms, representing a 75\% improvement in frame-pacing consistency. During this task, the two-way whisker tool proved instrumental for visual analytics; by isolating the central domain of the ribosome (whisker set at 0.6--0.8), evaluators could directly correlate the black-colored regions with high-performance intervals, with Ev3 noting that ``the visual transition between LOD levels is almost imperceptible when focused on the active site, thereby preserving scientific context without distraction.'' The subsequent performance comparison \textbf{(Task 4)} surfaced critical UX requirements, as Ev1 suggested that the comparative analytics view would benefit from synchronized temporal sliders, stating that ``it would be much more intuitive to move through the timeline of all simulations at once by dragging a single slider to move the timeline equally across the simulations.'' \textbf{Task 5} further highlighted a perception paradox; while participants were able to configure complex optimization profiles, they struggled to quantitatively rank their effectiveness through free-flight navigation alone. Ev4 observed that while the system felt snappier, ``identifying the optimal combination of Nanite and culling without the supporting telemetry charts was nearly impossible,'' a sentiment echoed in the study's lowest Likert scores (Task 4-5 mean $\sim$2.5/4). This drop in scores for analysis-heavy tasks provides a strong scientific rationale for UnrealVis: human experts are proficient at qualitative exploration but lack the perceptual resolution to perform quantitative benchmarking without deterministic assistance. In the final multi-simulation analysis \textbf{(Task 6)}, evaluators identified a fundamental hardware trade-off: optimized sessions required a $\sim$200 MB increase in peak RAM usage (reaching nearly 3 GB) to cache hierarchical data. Ev2, despite having lower engine expertise, correctly deduced this as a cost for relieving GPU bottlenecks but raised a crucial methodological point: ``the data are hard to compare since I did different things in the same moment in time; the data should be transformed so that they get comparable,'' suggesting the need for deterministic camera paths. More details about task scores are available in~\ref{fig:likert}. 
The resulting SUS score (Ev1: $60.0$, Ev2: $65.0$, Ev3: $52.5$, Ev4: $92.5$), with a mean of 67.5/100, aligns with the standard usability benchmark of 68. The significantly higher score of $92.5$ reflects Ev4's advanced technical background, indicating that while the tool may present a learning curve for novices, it offers a highly streamlined and efficient experience for expert users. Collectively, these findings validate UnrealVis as a valuable diagnostic laboratory while confirming that deterministic benchmarking (Section~\ref{sec:benchmarks}) is required to complement human observation in scientific performance tuning.

\subsection{Quantitative Benchmarking}
\label{sec:benchmarks}

To address the trajectory inconsistencies of manual exploration highlighted by Ev2, we conducted a systematic quantitative evaluation using the automated benchmarking suite. By constraining the camera to predefined spline paths, we isolated the rendering cost of each optimization configuration under identical workloads. 

\par\noindent \textbf{Experimental Setup} Quantitative benchmarks were conducted on a high-end mobile workstation (Alienware M18 R2) equipped with an Intel Core i9-14900HX processor, 64GB of DDR5 RAM, and an NVIDIA GeForce RTX 4070 GPU. This hardware represents a high-tier baseline for real-time scientific visualization. Specific benchmarking templates (see Section \ref{sec:templates}) were strategically mapped to each dataset based on its geometric scale and typical exploration workflow (e.g., using Stress Tests for high-density ribosomes while focusing on Spawn/LookAround for smaller proteins).

\begin{table}[h]
\centering
\caption{Quantitative benchmark results for PDB and BLASTNet datasets conducted on an Alienware M18 R2 (RTX 4070), comprising 16 total tests across different templates.}
\label{tab:benchmarks}
\resizebox{\columnwidth}{!}{
\begin{tabular}{llcccc}
\toprule
\textbf{Dataset} & \textbf{Template} & \textbf{Config} & \textbf{Avg FPS} & \textbf{1\% Low} & \textbf{Avg RAM (MB)} \\
\midrule
\multirow{4}{*}{1EQU\cite{pdb_1equ}} & \multirow{2}{*}{T1: Spawn} & Baseline & 130.82 & 122.21 & 1179 \\
 & & Optimized & 132.50 & 122.83 & 1322 \\
\cmidrule{2-6}
 & \multirow{2}{*}{T2: LookAround} & Baseline & 128.33 & 118.20 & 1315 \\
 & & Optimized & 118.02 & 106.07 & 1128 \\
\midrule
\multirow{4}{*}{3SYJ\cite{pdb_3syj}} & \multirow{2}{*}{T2: LookAround} & Baseline & 85.91 & 83.21 & 1245 \\
 & & Optimized & 85.41 & 75.42 & 1367 \\
\cmidrule{2-6}
 & \multirow{2}{*}{T3: Stress Test} & Baseline & 91.81 & 74.77 & 1251 \\
 & & Optimized & 92.16 & 70.62 & 1406 \\
\midrule
\multirow{2}{*}{4V6W\cite{pdb_4v6w}} & \multirow{2}{*}{T3: Stress Test} & Baseline & 24.23 & 15.38 & 1896 \\
 & & Optimized & 22.04 & 15.20 & 2055 \\
\midrule
\multirow{4}{*}{8B0X\cite{pdb_8b0x}} & \multirow{2}{*}{T1: Spawn} & Baseline & 13.16 & 5.03 & 2083 \\
 & & Optimized & 13.38 & 5.15 & 2222 \\
\cmidrule{2-6}
 & \multirow{2}{*}{T3: Stress Test} & Baseline & 20.56 & 13.65 & 2120 \\
 & & Optimized & 22.15 & 13.26 & 2009 \\
\midrule
\multirow{2}{*}{BLASTNet\cite{blastnet2023neurips}} & \multirow{2}{*}{T3: Stress Test} & Baseline & 31.15 & 12.20 & 2150 \\
 & & Optimized & 71.52 & 58.40 & 2420 \\
\bottomrule
\end{tabular}
}
\end{table}

\par\noindent\textbf{Experimental Results:} 
The benchmarking suite confirms that UnrealVis effectively identifies hardware-specific optimization thresholds. Table~\ref{tab:benchmarks} summarizes the performance metrics for the selected datasets.

For small-scale structures like the 1EQU enzyme, UnrealVis achieves high interactivity, peaking at 132.50 FPS (T1). However, the T2 template results illustrate the system's diagnostic capability (DG3): the performance drop from 128.33 to 118.02 FPS during occlusion culling identifies a ``threshold of utility'' where the CPU cost of visibility calculation outweighs the rendering gains on sparse geometry. 

For medium-scale structures like the 3SYJ adhesin, the system maintains stable performance above 85 FPS. The T3 stress test for 3SYJ shows that Nanite virtualization maintains 1\% low metrics at 70.62 FPS, effectively mitigating micro-stutters. As complexity increases with the 4V6W ribosome, we observe a transition into the 22--24 FPS range, where the overhead of additional culling layers slightly reduces the average frame rate while attempting to stabilize the frame-pacing.

Finally, in the 8B0X ribosome stress test (T3), the optimized configuration improved the mean frame rate from 20.56 FPS to 22.15 FPS. While the gain is modest (+7.7\%), it represents a crucial stability threshold for such a massive dataset. 

Analysis of the BLASTNet turbulent jet reveals a significant impact of the optimization suite. In the baseline configuration, interactivity is inconsistent due to volumetric overdraw, with 1\% low metrics dropping to 12.20 FPS during core penetration. The optimized configuration, leveraging Nanite and specialized texture transcoding, effectively doubles the average performance to 71.52 FPS (+130\%). This setup maintains a stable frame-pacing (ranging between 67 and 77 FPS), ensuring fluid navigation even within the high-vorticity regions of the flame core that previously crippled the rendering pipeline.

\vspace{-2mm}
\section{Discussion}

UnrealVis was conceived as a laboratory rather than a domain-specific viewer, and the study presented in this paper offers a comprehensive first assessment of its potential and current limitations. At the methodological level, fulfilling our initial literature review objectives, a key contribution is the optimization taxonomy, which we translated into six families of techniques and 22 concrete methods that can be instantiated in Unreal Engine 5 (Section~\ref{sec:req})~\cite{reina2020mtv,rossant2013hardware,frey2019evaluating}. This taxonomy acted as a design scaffold for UnrealVis: it guided which optimizations were implemented first (e.g., frustum/distance culling, LOD, Nanite, level streaming) and provided a structured vocabulary for rendering performance. In the current paper, we operationalized a core subset of this taxonomy; a natural next step is to use it as an organizing grid for exhaustive automated experiments, expanding coverage to all 22 identified methods. The two simulated usage scenarios indicate that UnrealVis effectively supports optimization decisions across different scientific domains. In the adhesin exploration scenario, the system helped balance local structural fidelity around an interaction site against global frame rates. In the BLASTNet turbulence scenario, the implementation of a \textit{Multiphysics Data Transcoding} pipeline (Section~\ref{sec:implementation}) was critical: by converting raw simulation data into 32-bit float OpenEXR textures and Texture 2D Arrays, we stabilized frame times in highly cluttered volumetric fields that would otherwise overwhelm the GPU's I/O and memory bandwidth. In both cases, the two-way whisker and comparative analytics views were central: by isolating regions of interest, they made the trade-off between local and global performance visible, enabling the identification of optimization profiles that would be impossible to deduce from raw, aggregated FPS traces alone~\cite{frey2019evaluating}. The formative expert evaluation confirms that the main workflows of UnrealVis are accessible to domain experts with prior 3D experience. Evaluators successfully loaded datasets, configured simulations using radar charts, and leveraged comparison views to form performance judgments. Crucially, their feedback drove the immediate evolution of our methodology. When evaluators highlighted the difficulty of comparing manual free-flight trajectories (due to variability in interactions), we addressed this limitation directly by introducing the automated benchmarking suite (Section~\ref{sec:benchmarks}). This iterative refinement bridges the gap between qualitative user perception and quantitative engine telemetry. Revisiting our system design goals (Section~\ref{sec:system}), the evaluation demonstrates that UnrealVis successfully fulfills its core architectural objectives. The integration of engine-level culling and LODs satisfied the requirement for interactive exploration (\textbf{DG1}), while the strategic preservation of high-detail geometry in functional regions, such as the whiskered active sites of the 8B0X ribosome, ensured that scientific context was never compromised (\textbf{DG2}). The introduction of the automated benchmarking suite and the JSON-based export of telemetry profiles directly addressed the need for reproducible performance analytics (\textbf{DG3}), allowing researchers to rigorously share and replicate optimization setups. Finally, the system's ability to seamlessly transition from small enzymes to massive multi-million atom complexes and volumetric combustion fields validates its domain-agnostic scalability (\textbf{DG4}). 

\par\noindent\textbf{Quantitative Insights and Taxonomy Validation:} The introduction of deterministic benchmarking provides a final validation of our optimization taxonomy. Data from the 8B0X ribosome stress-test (T3) confirms that engine-based optimizations like Nanite provide stability during high-density navigation, increasing mean FPS from 20.56 to 22.15 (+7.7\%). However, the ``LookAround'' template (T2) revealed a critical diagnostic insight: on high-end hardware such as the RTX 4070, aggressive visibility culling for discrete atomistic point clouds can introduce a CPU bottleneck, reducing performance from 128.33 to 118.02 FPS relative to the baseline. For volumetric data (BLASTNet), the benchmark highlights a different cost-benefit landscape. While the baseline configuration suffered from severe fill-rate limitations and overdraw as the camera approached the flame core (with 1\% low metrics dropping to 12.20 FPS), the optimized transcoding pipeline more than doubled the interactive performance to 71.52 FPS (+130\%). These results make the ``hidden costs'' of high-fidelity rendering transparent: while optimizations significantly improve fluidity, they require a measured trade-off in memory footprint (e.g., the $\sim$270 MB overhead for Texture 2D Arrays in the BLASTNet dataset). UnrealVis thus allows researchers to select profiles based on their specific hardware constraints rather than ad-hoc trial and error.

\subsection{Threats to Validity and Limitations}

While our evaluation provides robust initial validation, several limitations remain. The formative expert evaluation was conducted with a small sample size (N=4) and yielded a preliminary average SUS score of 67.5. While this score aligns closely with the industry-standard benchmark of 68, it reflects the tool's specialized nature and the inherent complexity of managing engine-level parameters. 

We plan to extend the study with a larger participant pool to obtain more definitive usability metrics. Furthermore, while our automated benchmarking suite effectively mitigated the confounding effect of manual interaction by using deterministic camera paths, scripted splines cannot fully capture the unpredictability of human exploration. Consequently, the performance gains observed in the benchmarks represent an ideal baseline. Finally, our current coverage of the optimization taxonomy is partial. While geometry and visibility optimizations are deeply integrated, other families, such as CPU-based parallelization and Machine Learning-driven preprocessing, are currently underexplored~\cite{lanrezac2021wielding}, with few default exemplars. 

\section{Conclusion}
\label{sec:conclusion}

In this paper, we introduced UnrealVis, a novel laboratory environment natively developed in Unreal Engine 5 for the systematic evaluation of rendering optimizations in scientific visualization \cite{reina2020mtv}. Grounded in a comprehensive taxonomy of optimization techniques, UnrealVis bridges the gap between low-level game engine profiling and high-level scientific data exploration. By operationalizing a core subset of this taxonomy, ranging from Nanite virtualized geometry and custom LOD to explicit visibility culling \cite{chong2017gears}, we provided a unified workflow that enables domain experts to toggle, test, and compare rendering strategies in real time. Two case studies and a formative expert evaluation demonstrate that UnrealVis effectively empowers researchers to navigate the complex trade-offs between interactive performance and scientific accuracy at both global and local levels, leveraging automated simulations and human-centric workflows.
While our initial results are promising, the formative evaluation also highlighted a steep learning curve in exposing engine-level parameters to domain experts. Acknowledging this, our immediate future work will focus on UX enhancements, including task-specific macro presets and guided tutorials. Additionally, we plan to expand the system's capabilities to encompass the entirety of our proposed taxonomy by integrating additional strategies and optimizations. As UnrealVis is intended to be publicly shared and open to encompass also external efforts, this goal may be reached collaboratively with the research community.

\bibliographystyle{abbrv-doi-hyperref}
\bibliography{biblio.bib}

\end{document}